\documentclass[ reprint, amsmath, amssymb, aps, pra ]{revtex4-2}

\usepackage{graphicx}
\usepackage{bm}
\usepackage{amsmath,amssymb}
\usepackage{color}
\usepackage[caption=false]{subfig}
\usepackage{enumitem}

\usepackage{dcolumn}
\usepackage{hyperref}
\usepackage[mathlines]{lineno}

\usepackage{gensymb} 
\usepackage{makecell} 
\usepackage[normalem]{ulem} 
\usepackage{float} 
\usepackage{threeparttable} 

\usepackage{array}
\newcolumntype{P}[1]{>{\centering\arraybackslash}p{#1}}

\newcommand{\gh}{{g^{(2)}_\mathrm{h}}}

\newcommand{\Rsync}{{R_\mathrm{sync}}}
\newcommand{\Rstoc}{{R_\mathrm{stoc}}}
\newcommand{\eff}{{\eta_\mathrm{e2e}}}
\newcommand{\effbar}{{\bar{\eta}_\mathrm{e2e}}}

\newcommand{\beginsupplement}{%
	\setcounter{table}{0}
	\renewcommand{\thetable}{S\arabic{table}}%
	\setcounter{figure}{0}
	\renewcommand{\thefigure}{S\arabic{figure}}%
	\setcounter{equation}{0}
	\renewcommand{\theequation}{S\arabic{equation}}%
	\setcounter{section}{0}
	\renewcommand{\thesection}{S\arabic{section}}%
}

\hypersetup{
    colorlinks,
    citecolor=black,
    filecolor=black,
    linkcolor=black,
    urlcolor=black
}

\begin{document}

\title{Single-photon synchronization with a room-temperature atomic quantum memory}

\author{Omri Davidson}
\affiliation{Department of Physics of Complex Systems, Weizmann Institute of Science, Rehovot 7610001, Israel}
\author{Ohad Yogev}
\affiliation{Department of Physics of Complex Systems, Weizmann Institute of Science, Rehovot 7610001, Israel}
\author{Eilon Poem}
\affiliation{Department of Physics of Complex Systems, Weizmann Institute of Science, Rehovot 7610001, Israel}
\author{Ofer Firstenberg}
\affiliation{Department of Physics of Complex Systems, Weizmann Institute of Science, Rehovot 7610001, Israel}

\begin{abstract}
\noindent
Efficient synchronization of single photons that are compatible with narrowband atomic transitions is an outstanding challenge, which could prove essential for photonic quantum information processing. 
Here we report on the synchronization of independently-generated single photons using a room-temperature atomic quantum memory. The photon source and the memory are interconnected by fibers and employ the same ladder-level atomic scheme. We store and retrieve the heralded single photons with end-to-end efficiency of $\eff=25\%$ and final anti-bunching of $\gh=0.023$. Our synchronization process results in over tenfold increase in the photon-pair coincidence rate, reaching a rate of more than $1000$ detected synchronized photon pairs per second. The indistinguishability of the synchronized photons is verified by a Hong-Ou-Mandel interference measurement.
\end{abstract}

\maketitle

Multi-photon states are an important resource for photonic quantum information processing, with potential applications in quantum computation, communication, and metrology \cite{Photonics_QIP_review_2019, 2001_QC_using_linear_optics, 2015_multiphoton_quantum_communication, Quantum_metrology_with_photons_2008_Dowling}. It is beneficial that these photons interact coherently with atomic ensembles, to enable the implementation of deterministic two-photon gates \cite{2016_Firstenberg_Rydberg_review} and quantum repeaters for long-distance communication \cite{2001_DLCZ_paper}. 
Efficient, well-established, room-temperature platforms for generating such photons are based on 
parametric processes such as spontaneous parametric down-conversion (SPDC) and four-wave mixing (FWM) \cite{Eisman_2011_review_single_photons}. These processes give rise to stochastic emission of photon pairs and are therefore utilized as heralded single-photon sources \cite{2016_SPDC_in_cavity_Laurat,2017_photon_source_cavity_SPDC_Chuu,2020_source_FWM_in_MRR_Thompson,photon_source_CW_cold_Kurtriefer_2014}. However, they are probabilistic, rendering the construction of larger multi-photon states exponentially slow \cite{2013_Nunn_enhancing_rates_with_memories}. At present, the demonstrated rate for generating a 12-photon entangled state from six stochastic emission events is one per hour \cite{2018_Pan_SPDC_12_photon_entanglement}. 

The exponential scaling of the rate with the number of photons $N$ can be mitigated by using quantum memories to synchronize the probabilistically generated photons \cite{2013_Nunn_enhancing_rates_with_memories}. Particularly, the quantum memory can support a time-multiplexing scheme,  generating a string of quasi-deterministic photons at pre-determined clock cycles \cite{2002_Pittman_storage_loop,2015_Kwiat_storage_loop,2019_Kwiat_storage_loop}. Alternatively, $N$ stochastic photon sources with $N-1$ memories can be used to generate a synchronous $N$-photon state \cite{2013_Nunn_enhancing_rates_with_memories, 2017_Kwiat_storage_loop}. 
Most works focus on $N=2$, and we do so as well. 

For $N=2$, we identify several key metrics. The first is the rate enhancement factor $\zeta=\Rsync/\Rstoc$, which is the accomplishment of the synchronization, the ratio between the detection rate of photon pairs after the synchronization $\Rsync$ compared to the stochastic (accidental) pair detection rate before synchronization $\Rstoc$. The second is $\Rsync$, which should be high for practical applications. A third metric is the anti-bunching of the synchronized photons $\gh$, which is the normalized autocorrelation of the retrieved signal field conditioned on heralding. Ideally, $\gh=0$, and any undesired multi-photon component, \textit{e.g.}, due to noise, increases $\gh$.   

There are two types of memories: those containing an internal source \cite{photon_soure_DLCZ_cold_Pan_2006, photon_soure_DLCZ_cold_Granjier_2014, 2021_Poltzik_DLCZ_source}, and input-output memories accepting photons from outside \cite{2019_Du_cold_memory, 2020_Laurat_memory_for_single_photons_cold_atoms, 2022_Treutlein_vapor_memory}. The natural advantage of input-output memories is that they can be used to synchronize and delay photons that have already undergone some processing, including photons that are part of larger entangled states. 
Photon synchronization has been demonstrated with cold \cite{2006_Kimble_photon_synchronization_DLCZ, 2007_Pan_photon_synchronization_DLCZ, 2021_Duan_Photon_synchronization_DLCZ, 2021_Zhu_photon_synchronization_cold_atoms} and hot \cite{2022_Xian_Min_Photon_synchronization_DLCZ_hot} atomic ensembles, employing internal-source memories \cite{2006_Kimble_photon_synchronization_DLCZ, 2007_Pan_photon_synchronization_DLCZ, 2021_Duan_Photon_synchronization_DLCZ, 2022_Xian_Min_Photon_synchronization_DLCZ_hot} and input-output memories \cite{2021_Zhu_photon_synchronization_cold_atoms}. 
However, all these demonstrations suffer from a low photon-pair synchronization rate [$\Rsync<1$ counts per second (cps)] and moderate photon antibunching ($\gh>0.15$). For a comparison of different experiments, see Supplemental Material (SM). 

A successful, competing approach to atomic memories uses all-optical setups, namely cavities 
 \cite{2013_Furusawa_concatenated_cavities_source_and_memory, 2016_Furusawa_synchronization} and storage loops
\cite{2002_Pittman_storage_loop,2015_Kwiat_storage_loop, 2017_Kwiat_storage_loop, 2019_Kwiat_storage_loop, 2020_Xian_Min_Jin_synchronization, 2022_Silberhorn_storage_loop,2023_Guo_synchronization_storage_loop}. 
Cavity systems have achieved good performance with narrowband photons, $\zeta=25$ and $\Rsync=90$ cps \cite{2016_Furusawa_synchronization}, but these are internal-source systems.
Storage loops, which are input-output systems,  have reached $\zeta=30$ and $\Rsync=450$ cps with broadband SPDC photons \cite{2017_Kwiat_storage_loop, 2019_Kwiat_storage_loop, 2022_Silberhorn_storage_loop} but inferior performance with narrowband photons \cite{2020_Xian_Min_Jin_synchronization}.
Notably, interfacing the broadband photons generated from SPDC with atomic ensembles remains an outstanding challenge.

Here we demonstrate for the first time single-photon synchronization using an input-output memory that combines substantial rate enhancement $10\le \zeta \le 30$, high pair detection rates $\Rsync\le1200$ cps, low-noise operation with $\gh=0.023$, and compatibility with atomic ensembles. 
We achieve these at room temperature by employing the ladder orbital scheme $|5S_{1/2}\rangle \rightarrow |5P_{3/2}\rangle \rightarrow |5D_{5/2}\rangle$ in rubidium vapor for the photon source \cite{photon_source_CW_hot_SebMoon_2016, Photon_source_paper, Photon_source_paper_2} \textit{and} for the quantum memory \cite{FLAME_paper,FLAME_2_paper}. 
This scheme has three main benefits. First, the all-orbital fast ladder memory (FLAME) provides high-bandwidth operation, low noise, and high end-to-end memory efficiency \cite{ORCA_ladder_memory,FLAME_paper,FLAME_2_paper} which are key for high-rate single-photon synchronization. Second, the small wavelength mismatch within the two-photon transition enables a nearly Doppler-free operation and thus a long coherence time between the ground and doubly-excited state. This provides a memory lifetime of over $100$ ns \cite{FLAME_2_paper} and single-photon generation with high rate and low noise \cite{photon_source_CW_hot_SebMoon_2016, Photon_source_paper, Photon_source_paper_2}. Third, by employing the same level scheme for the photon source and quantum memory, the generated photons are inherently compatible with the memory, enabling an end-to-end memory efficiency of $\eff=25\%$. This also maintains the indistinguishability of the synchronized photons, as quantified by the Hong-Ou-Mandel (HOM) interference visibility $V_\text{sync}=76\%$.

\begin{figure} 
	\centering
	\includegraphics[width=\columnwidth,trim=8.0cm 2.0cm 4.0cm 0.5cm ,clip=true] {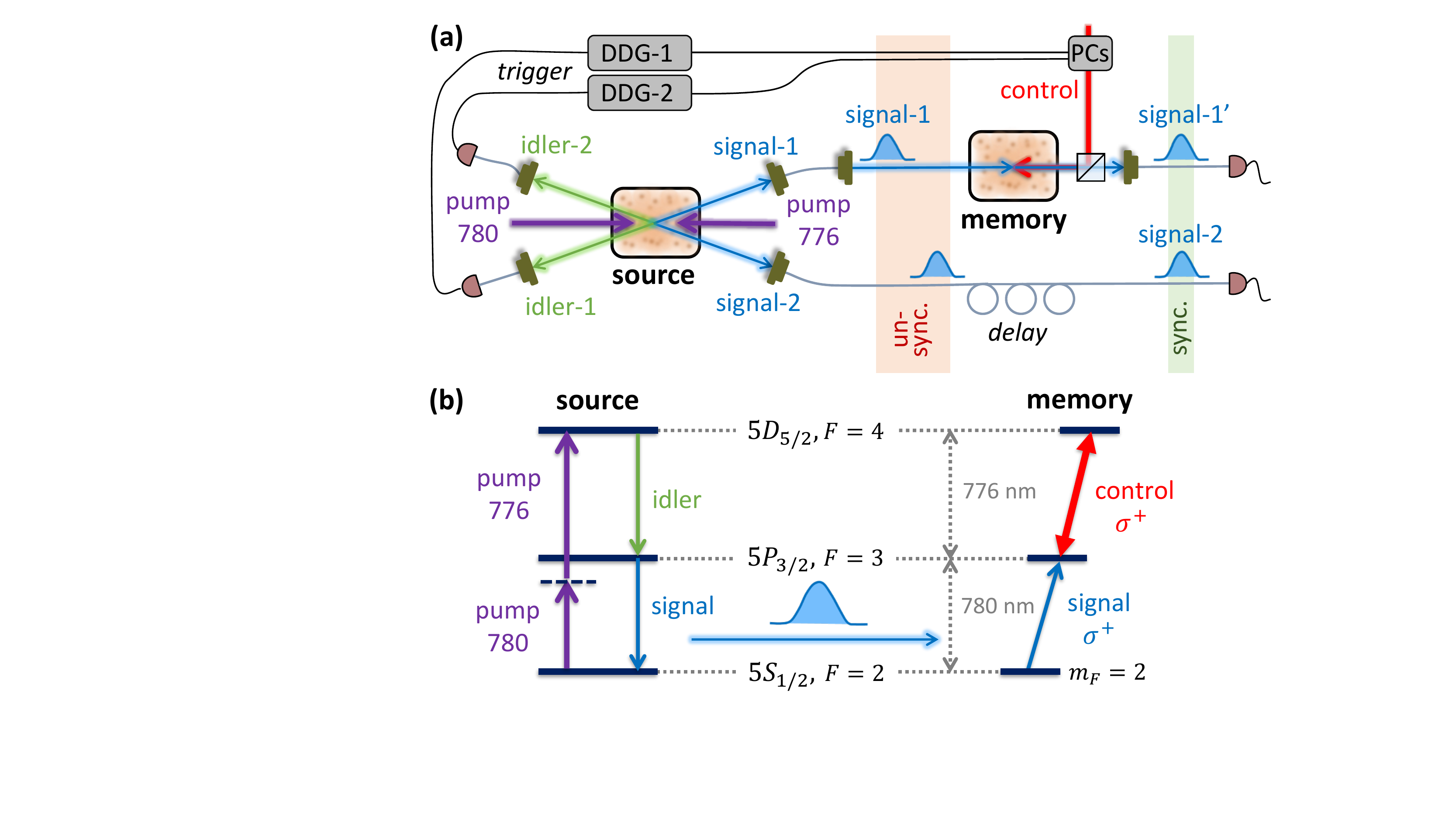}
	\caption{\textbf{Photon synchronization scheme.} 
	(\textbf{a}) Sketch of the experimental setup. Two pump fields continuously excite the atoms in the source module, which then emit signal and idler photons in two phase-matched directions via four-wave mixing. Signal-1 photon in the first collection channel goes to the memory module, while the detection of idler-1 triggers the control storage-pulse in the memory [generated by Pockels cells (PCs)]. Signal-2 in the second collection channel goes into a fiber delay line, while the detection of idler-2 triggers the control retrieval-pulse, which releases signal-1$'$ from the memory synchronously with signal-2. 
    (\textbf{b}) The photon source and memory both employ the same ladder-type level system of $^{87}\text{Rb}$, which is nearly Doppler-free and enables high storage efficiency and fidelity. 
	}
	\label{fig:experiment} 
\end{figure}

\emph{Synchronization scheme.}---
The synchronization experiment comprises a spatially-multiplexed single-photon source, a quantum memory, and electronic triggering of the memory operation, as shown schematically in Fig.~\ref{fig:experiment}. 
The photon source is based on FWM in rubidium vapor with two continuous-wave pump fields \cite{Photon_source_paper,Photon_source_paper_2}. The pump fields, at wavelengths of 780~nm and 776~nm, counter-propagate through an isotopically purified $^{87}\text{Rb}$ vapor cell and excite the $|5S_{1/2},F=2\rangle \rightarrow |5P_{3/2},F=3\rangle$ and $|5P_{3/2},F=3\rangle \rightarrow |5D_{5/2},F=4\rangle$ transitions, respectively. The detection of a spontaneously emitted idler photon heralds the generation of a collective state comprising a single $|5P_{3/2}\rangle$ excitation that is shared among all atoms, and the signal photon emission to the ground state is thus collectively enhanced into the phase-matched direction \cite{Jen_2012_superradiance_in_cascade_systems}. We utilize the cylindrical symmetry of the phase-matching condition to set collection channels on both sides of the optical axis, effectively realizing two sources in the same vapor cell. We denote the generated photons in channel $j$ as idler-$j$ and signal-$j$ ($j=1,2$). Additional details on the photon source are given in SM and in Refs.~\cite{Photon_source_paper, Photon_source_paper_2}. 

The quantum memory is based on the FLAME scheme in $^{87}\text{Rb}$ vapor \cite{FLAME_paper,FLAME_2_paper}. Initially, the atoms in the memory cell are optically pumped to the maximal spin state. An input signal-1 photon, which couples to the $|5S_{1/2},F=2,m_F=2\rangle \rightarrow |5P_{3/2},F=3,m_F=3\rangle$ transition, is stored on the doubly-excited state $|5D_{5/2},F=4,m_F=4\rangle$ by sending the first (storage) control pulse. At a controllable time later, a second (retrieval) control pulse releases the signal-1$'$ photon from the memory (1$'$ marks post-memory). We use an auxiliary dressing field (not shown in Fig.~\ref{fig:experiment}) that weakly couples the storage state $|5D_{5/2}\rangle$ to a high-lying orbital in order to counteract the residual Doppler broadening of the two-photon transition \cite{Continuous_protection_paper} and prolong the memory lifetime \cite{FLAME_2_paper}. The overall transmission of the memory module from the input fiber to the output fiber is $T=68\pm2\%$ (including the 4\% loss on exiting the input fiber). Further details on the memory are given in SM and in Ref.~\cite{FLAME_2_paper}.

The detection of idler photons triggers digital delay generators (DDG) that set off the control pulses for the memory via free-space Pockels cells (PCs). DDG-1, triggered by idler-1, sends a control pulse that stores the heralded signal-1 in the memory. Subsequently, DDG-2, triggered by idler-2, sends a second control pulse that retrieves signal-1$'$. This protocol synchronizes signal-1$'$ and signal-2.
We find that the memory efficiency is optimal when the control field is on resonance, indicating that signal-1 is emitted from the source on resonance, as expected \cite{Photon_source_paper_2}.
As our PCs' maximal average repetition rate is limited to $3\times 10^5$ operations per second, we devise a logical scheme that operates them only if idler-1 and idler-2 were both detected within a 100-ns time window, set by the memory lifetime. Details on the electronic triggering, timing sequence, and fiber routing are given in SM.


\begin{figure*} 
	\centering
	\includegraphics[width=\textwidth,trim=0.0cm 0.0cm 0.0cm 0.0cm ,clip=true] {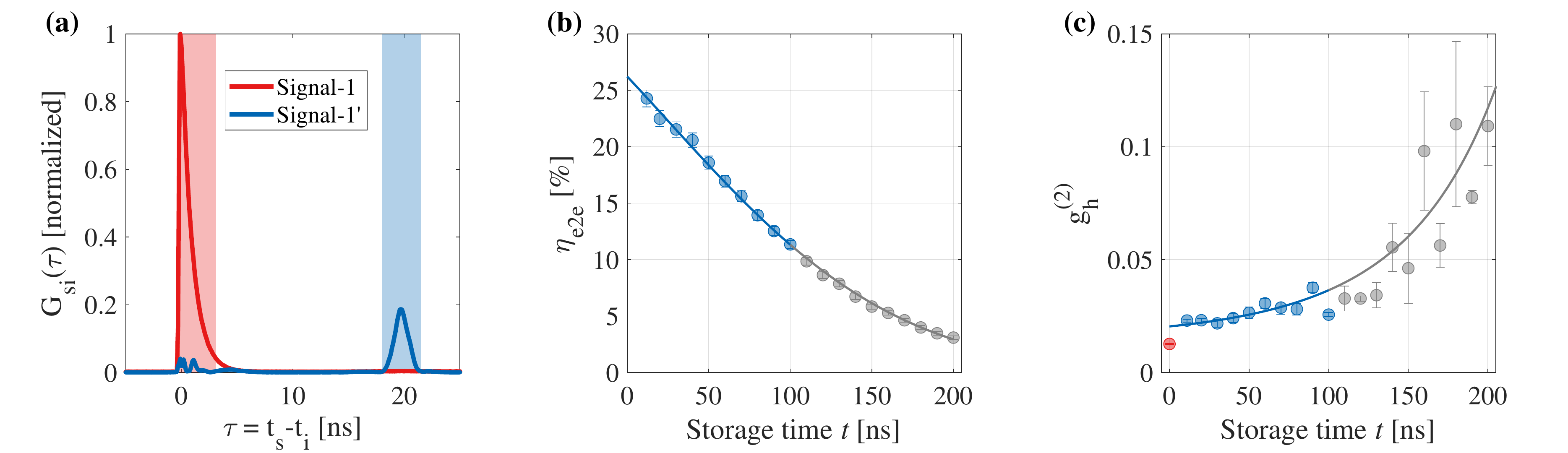}
	\caption{\textbf{Storage of heralded single photons.} 
	(\textbf{a}) Raw histogram counts (signal-idler cross-correlation) without storage (signal-1, red) and after storage and retrieval in the memory (signal-1$'$, blue). Here $\tau=t_\text{s}-t_\text{i}$ is the time difference between detections of the signal and idler photons. The shaded areas indicate the 3.5-ns-long integration window used throughout the paper. 
    (\textbf{b}) End-to-end memory efficiency versus storage time of heralded single photons. Circles are the measured data,  and the line is a fit to a model of exponential and Gaussian decays. The errorbars comprise the standard error of repeated measurements and the uncertainty on the detection efficiencies of signal-1 and signal-1$'$ (see SM). 
    (\textbf{c}) The normalized auto-correlation of signal-1$'$ conditioned on the detection of idler-1, $\gh$, indicating the multi-photon component in the retrieved field. The $\gh$ of un-stored photons (signal-1) is shown in red for reference. The line is a fit to a model comprising the finite memory efficiency and assuming that noise photons originate only from the source.
    In (\textbf{b}) and (\textbf{c}), blue indicates the range of storage times $t\le 100$ ns used in the synchronization experiment. 
	}
	\label{fig:Quantum storage} 
\end{figure*}

\emph{Storage and retrieval of heralded single photons.}---
We begin by characterizing the storage and retrieval of signal-1.  
Figure~\ref{fig:Quantum storage}(a) shows the count histogram [signal-idler correlation $G_\text{si}(\tau)$] for a storage time of $t=20$~ns. We compare the histogram of signal-1 (\textit{i.e.}~directly after the photon source) to that of signal-1$'$ (\textit{i.e.}~after storage and retrieval in the memory, including the overall transmission of the memory module). 
The 3.5-ns-long shaded areas indicate the integration windows used for calculating $\eff$, $\gh$, $\Rstoc$, $\Rsync$, and $V_\text{sync}$; This window captures over $95\%$ of the pulse energy. 

The memory efficiency $\eff$ versus the storage time $t$ is shown in Fig.~\ref{fig:Quantum storage}(b). Here, we directly measure the end-to-end efficiency by connecting the optical fiber of signal-1 either to the memory input fiber ($98\pm1\%$ coupling) or directly to the detector input fiber ($92\pm1\%$ coupling). Comparing between the detection rates of signal-1 and signal-1$'$, after correcting for the different couplings, provides $\eff$. Note particularly that $\eff$ includes all fiber/free-space couplings.
We measure $\eff(t=12\text{ ns})=24.3\pm0.8 \%$. By fitting the data to a decoherence model $\eta(t)=\eta(0)e^{-t^2/2\tau_\sigma^2-t/\tau_\gamma}$ with homogeneous ($\tau_\gamma$) and inhomogeneous ($\tau_\sigma$) decoherence times, we extract the zero-time memory efficiency $\eff(0)=26.2\pm0.5 \%$. The memory $1/e$ lifetime is $\tau_s=114\pm 2$ ns. Here the errors are 1 standard deviation (s.d.)~of the fit uncertainty. 
Given the overall transmission $T$, the memory internal efficiency, comprising only the mapping of the light to and from the atoms, is $\eta_\text{int}(0)=38.4\pm 1.1 \%$. 

We verify that the memory preserves the quantum statistics $\gh\ll 1$ of the stored single photons, as shown in Fig.~\ref{fig:Quantum storage}(c). For $t=20$~ns, the multi-photon component of signal-1$'$ is $\gh=0.023\pm0.001$, which is higher than $\gh=0.0126\pm0.0002$ of signal-1 but still at the few-percent level.

The increase in $\gh$ is due to noise photons originating in either the memory or the source. In our system, the former is negligible: the memory generates only $\nu = (1.7\pm0.2)\times 10^{-5}$ noise photons per operation. These photons govern the short-time signal-to-noise ratio $\text{SNR}=\eta_\text{h}\eff(0)/\nu =3100\pm 400$, where $\eta_\text{h}=20\%$ is the heralding efficiency of the source, and indeed $\text{SNR}^{-1}\ll \gh$. Therefore, we attribute the increase in $\gh$ predominantly to noise photons arriving from the source at the time of retrieval and detected in coincidence with signal-1$'$. The dominant contribution comes from photons that scatter directly from the continuous, off-resonant, 780-nm pump field, which are transmitted well through the memory module. Further increase of $\gh(t)$ at larger $t$ is explained solely by the decrease of $\eff(t)$.
Our model of this mechanism, shown in Fig.~\ref{fig:Quantum storage}(c) and detailed in SM, agrees well with the data.

\begin{figure} 
	\centering
	\includegraphics[width=0.95\columnwidth,trim=0.0cm 3.0cm 0.5cm 0.0cm ,clip=true] {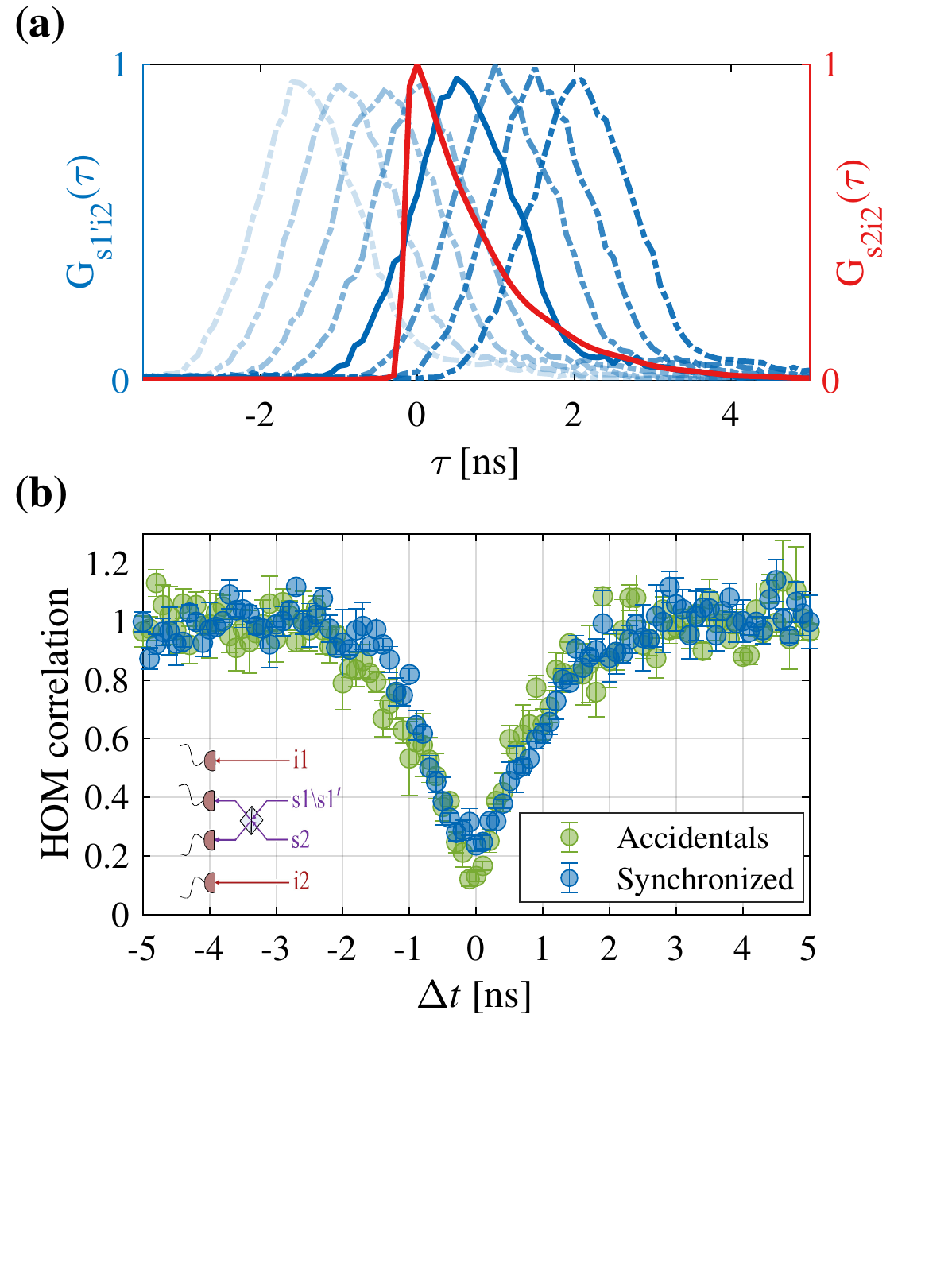}
    \caption{\textbf{Photon synchronization.}
    (\textbf{a}) Histograms of signal-2-idler-2 correlation (red) and signal-1$'$-idler-2 correlation (blue), demonstrating the synchronization of signal-1$'$ with signal-2. The curves with different shades of blue correspond to different controlled retrieval times of signal-1$'$ with 500 ps intervals. The solid-blue curve corresponds to the synchronized signal-1$'$. 
    (\textbf{b}) HOM interference between photons originating from the two source channels. Green: signal-1-signal-2 (without the memory), where $\Delta t$ is the time difference between idler-1 and idler-2. Blue: signal-1$'$-signal-2 (with the memory and synchronization), where $\Delta t$ is controlled by tuning the electronic delay between idler-2 (trigger) and signal-1$'$ (memory retrieval). The inset shows a schematic of the detection scheme.
	}
	\label{fig:synch and HOM} 
\end{figure}

\emph{Photon synchronization.}---
We now turn to demonstrate the synchronization of photon pairs using the memory. Figure~\ref{fig:synch and HOM}(a) shows temporal profiles of the retrieved signal-1$'$ photons (signal-1$'$-idler-2 correlation conditioned on memory operation, $G_{\text{s}{1'}\text{i}2}$) in comparison to the profile of
signal-2 (signal-2-idler-2 correlation, $G_\text{s2i2}$) for varying timing settings. Note that the data do not correspond to a specific memory time $t$ but rather represent an average over $0<t\le100$ ns, stochastically `sampled' by the regular operation of the synchronization protocol. We control the exact relative timing $\Delta t$ between signal-1$'$ and signal-2 by electronically tuning the trigger delay, which controls the memory retrieval time. Figure~\ref{fig:synch and HOM}(a) demonstrates, for arbitrary 500-ps intervals, our capability of on-demand, continuous tuning of the retrieval time.

To optimize the relative timing and to characterize the fidelity of the synchronized photons, we perform HOM interference measurements \cite{HOM_1987}. These measurements attest to the indistinguishability of the synchronized photon pair.
Figure~\ref{fig:synch and HOM}(b) shows the HOM correlation of signal-1$'$ and signal-2 for varying triggering delays. The HOM visibility (the interference contrast), quantifying the indistinguishability, is $V_\text{sync}=76\pm2 \%$. We use the position of the minimum to define $\Delta t=0$ and to set the optimal delays in the system. 
For reference, we show in Fig.~\ref{fig:synch and HOM}(b) the HOM measurement of accidental pairs without the memory, exhibiting $V_\text{stoc}=88\pm2$\%. 
Notably, the acquisition time per data point for synchronized pairs is 100 times shorter than for accidental pairs, illustrating the importance of synchronization for efficiently manipulating multi-photon states.

\begin{figure} 
	\centering
	\includegraphics[width=0.95\columnwidth,trim=0.0cm 1.5cm 0.0cm 0.0cm ,clip=true] {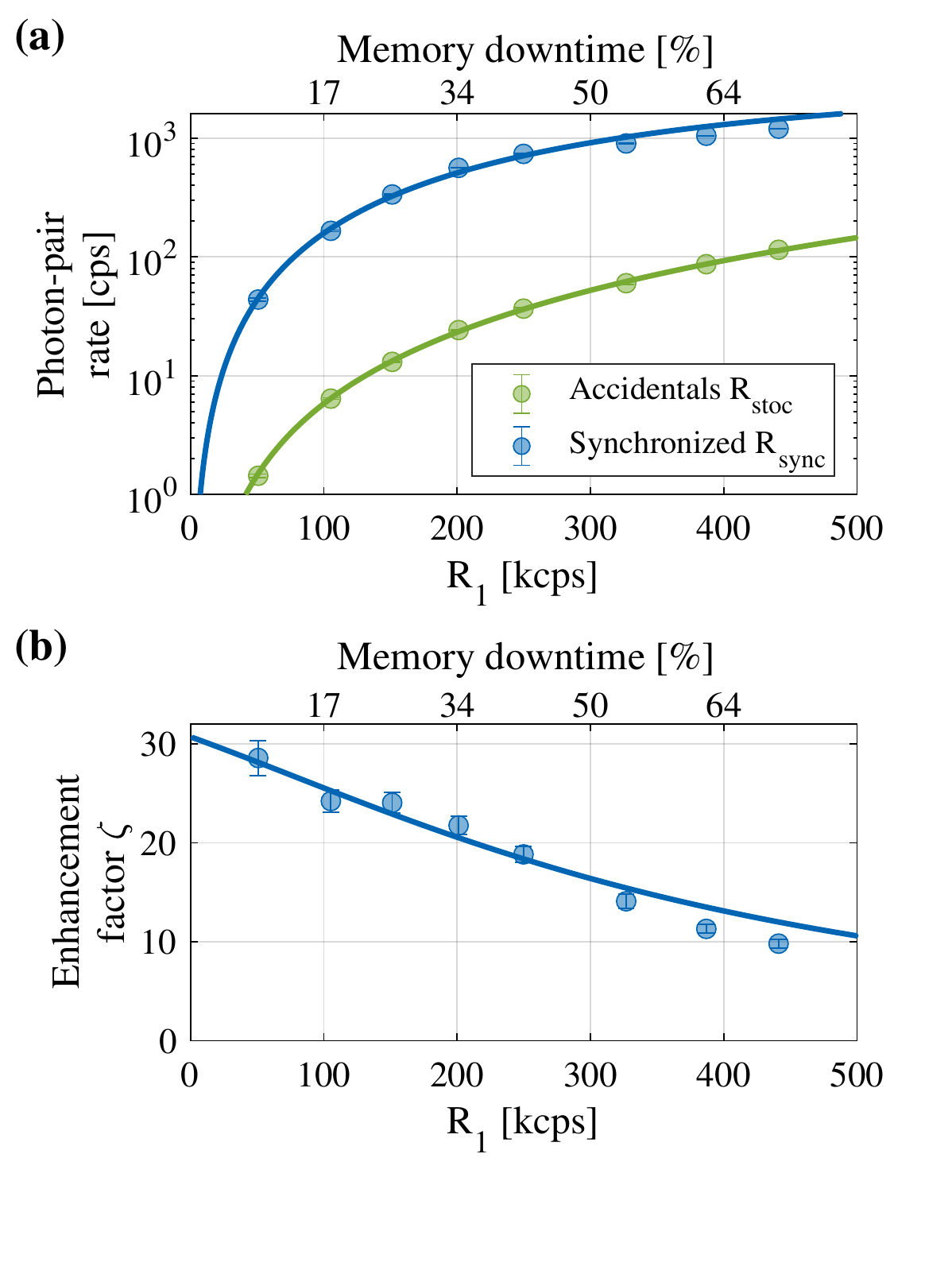}
	\caption{\textbf{Rate enhancement by synchronization.}
    (\textbf{a}) Pair coincidence count rate with the memory ($\Rsync$, blue) and without the memory ($\Rstoc$, green) versus the single-photon count rate $R_1$. Accidental coincidences from the source are considered if the photons arrive within $\pm 300$ ps from one another (see text). (\textbf{b}) The enhancement factor $\zeta=\Rsync/\Rstoc$ of pair coincidence rate.
    Circles are measured data, and the lines are calculations based on independently-measured parameters of the source and memory with no fit parameters. The top-horizontal axes show the relative downtown of the memory, during which it cannot handle photons for synchronization due to technical limitations. This downtown quantifies the technical saturation of the system, responsible for the degradation of $\zeta$ with $R_1$.
    }
	\label{fig:Rate enhancement} 
\end{figure}

Finally, we show the merit of synchronization in terms of the enhanced two-photon coincidence rate $\Rsync$ in Fig.~\ref{fig:Rate enhancement}(a). We study $\Rsync$ as a function of the heralded single-photon rate $R_1$, which we control by tuning the strength of the pumps in the source module. We cover the range $50<R_1<440$ kilo-cps (kcps), for which the accidental coincidence rates are $1.4<\Rstoc<115$ cps. Here, we consider an accidental coincidence if idler-1 and idler-2 are detected within $\pm 300$ ps of each other.  
 This time interval is chosen based on the HOM correlation measurement, which exhibits $V_\text{stoc}\ge 75\%$ within $\pm 300$ ps around the minimum (this is a conservative choice, generous in terms of $\Rstoc$).
As shown in Fig.~\ref{fig:Rate enhancement}(a), the coincidence rates after synchronization grow to $44<\Rsync<1200\pm10$ cps, a substantial enhancement compared to $\Rstoc$.

The rate enhancement $\zeta=\Rsync/\Rstoc$ is shown in Fig.~\ref{fig:Rate enhancement}(b). The maximal enhancement $\zeta=28.6\pm1.8$ is obtained at low $R_1$. As $R_1$ increases, the system is triggered more often, and each triggering event is followed by $\sim 1~\mu$s during which the memory cannot handle additional photons (see SM for details). This leads to technical saturation of the system, which we quantify by the relative memory downtime, shown as top axes in Fig.~\ref{fig:Rate enhancement}.
A second issue occurring at high $R_1$ is a moderate increase of $\gh$ and a corresponding decrease of $V_\text{sync}$ (see SM). We attribute this partially to a degradation of the pulses generated by the PCs at a high triggering rate.
Nevertheless, even at the highest rate, we obtain a tenfold increase in the pair coincidence rate and a non-classical HOM visibility $V_\text{sync}>50\%$.

The rates of accidental and synchronized photon pairs and their dependence on the rate of single photons can be calculated from the parameters of the source, memory, and electronics, all of which we have independently characterized. Our model, based solely on these parameters without fitting (see SM for details), is presented by the solid lines in Fig.~\ref{fig:Rate enhancement}. The model correctly predicts $\Rsync$ and $\Rstoc$ and confirms that the decrease of $\zeta$ with $R_1$ is due only to the memory downtime. We attribute the slight deviation of the model from the data at large $R_1$ to the PCs' pulse degradation noted above.

\emph{Discussion.}---
There are several factors limiting the increase of the two-photon coincidence rate. The main ones are the end-to-end efficiency of the memory, the limited operation rate of the PCs, and the heralding efficiency of the photon source. All these factors can be improved. 

First, as discussed in Ref.~\cite{FLAME_2_paper}, practicable technical improvements of the memory module can increase the internal memory efficiency to $65$\%. This, in addition to raising the setup transmission by anti-reflection coating of the optical fiber-to-free-space interfaces, will substantially raise the end-to-end efficiency. 
Second, the heralding efficiency of the photon source can be increased by using etalon filters to block the direct scattering of photons from the pump fields into the idler modes.
Third, the PCs can be replaced by an amplitude electro-optic modulator seeding a tapered amplifier \cite{2022_Treutlein_vapor_memory}. This will enable both a higher repetition rate and a higher memory efficiency by optimizing the control temporal shape to that of the signal photons \cite{Gorshkov_optimal_storage_letter,Gorshkov_free_space_model}. 

In conclusion, we demonstrate synchronization of single photons with a high rate and low noise using a quantum memory and a photon source, both based on a ladder-level scheme in rubidium vapor. Our synchronized photons are well-suited for quantum information protocols requiring efficient interaction with atomic ensembles, such as Rydberg-mediated deterministic two-photon gates. The scheme presented here can be used to efficiently generate synchronous few-photon states, and, with feasible improvements, larger multi-photon states.

We acknowledge financial support from the Israel Science Foundation, the US-Israel Binational Science Foundation (BSF) and US National Science Foundation (NSF), the Minerva Foundation with funding from the Federal German Ministry for Education and Research, the Estate of Louise Yasgour, and the Laboratory in Memory of Leon and Blacky Broder.


\bibliographystyle{unsrt}

\bibliography{Photon_synchronization_bibliography.bib}

\clearpage

\beginsupplement
\onecolumngrid

\begin{center} \Large \textbf{Supplementary material} \Large \end{center}


The supplementary material contains the following sections:
\begin{itemize}
    \item \ref{synchronization comparison}. Comparison of photon synchronization experiments

    \item \ref{experimental setup}. Experimental setup

    \begin{itemize}
        \item \ref{photon source setup}. Photon source
        \item \ref{memory setup}. Memory
        \item \ref{electronics}. Electronics and timing
        \item \ref{optical fibers}. Optical fibers and photon detection
    \end{itemize}

    \item \ref{coincidence rate model}. Model for photon coincidence rate

    \item \ref{g2 model}. Model for $\gh$

    \item \ref{data analysis}. Data analysis

    \item \ref{Additional experimental data}. Additional experimental data
    
\end{itemize}

\section{Comparison of photon synchronization experiments} \label{synchronization comparison}
Table~\ref{tab:synchronization comparison} compares different photon-synchronization experiments. We only consider experiments that perform active photon-synchronization with a quantum memory. 

\begin{table}[H] 
\caption{
\textbf{Comparison of photon-synchronization experiments.} 
We compare the detected photon pairs after synchronization $\Rsync$ in counts per second (cps), the enhancement factor of the memory $\zeta$, the multi-photon component of the synchronized photons $\gh$, and the measured HOM interference visibility. Note that we show the best reported values in each metric, not all of which are simultaneously achieved.
}
\label{tab:synchronization comparison}
\vspace{0.2cm}
\centering
\begin{threeparttable}
\begin{tabular}{|P{0.8cm}|c|c|P{3.0cm}|P{2.5cm}|c|P{1.7cm}|}
\hline
\textbf{Ref.} &
  \textbf{Source} &
  \textbf{Memory} &
  \textbf{Synchronization rate} $\mathbf{R}_\textbf{sync}$ \textbf{[cps]} &
  \textbf{Enhancement factor} $\mathbf{\zeta}$  &
  $\mathbf{g}_\textbf{h}^{\mathbf{(2)}}$ &
  \textbf{HOM visibility [\%]} \\ \hline
This work                                             & Hot atoms  & Hot atoms                 & 1200   & 28.6  & 0.023 & 76 \\ \hline
\cite{2006_Kimble_photon_synchronization_DLCZ}       & Cold atoms & Cold atoms$^\text{a}$              & 0.03   & 28    & 0.17  & 77   \\ \hline
\cite{2007_Pan_photon_synchronization_DLCZ}          & Cold atoms & Cold atoms$^\text{a}$              & 0.08   & 136$^\text{b}$ & 0.17  & 80   \\ \hline
\cite{2021_Duan_Photon_synchronization_DLCZ}         & Cold atoms & Cold atoms$^\text{a}$              & 0.167  & 300$^\text{b}$ & -     & -    \\ \hline
\cite{2021_Zhu_photon_synchronization_cold_atoms}    & Cold atoms & Cold atoms                & 0.89   & 15    & 0.43  & 76   \\ \hline
\cite{2022_Xian_Min_Photon_synchronization_DLCZ_hot} & Hot atoms  & Hot atoms$^\text{a}$               & 0.0056 & 15    & 0.33  & 75   \\ \hline
\cite{2016_Furusawa_synchronization}                 & SPDC       & Cavity$^\text{a}$ & 90     & 25$^\text{b}$  & -     & 82   \\ \hline
\cite{2020_Xian_Min_Jin_synchronization}             & Hot atoms  & Storage loop              & 1.3$^\text{c}$ & -     & -     & -    \\ \hline
\cite{2017_Kwiat_storage_loop}                       & SPDC       & Storage loop              & 14     & 30.5  & -     & 91$^\text{d}$ \\ \hline
\cite{2022_Silberhorn_storage_loop}                  & SPDC       & Storage loop              & 450    & 9.7   & -     & 94.5 \\ \hline
\cite{2023_Guo_synchronization_storage_loop}         & SPDC       & Storage loop              & 6    & 7.5   & -     & 91.7 \\ \hline
\cite{2019_Kwiat_storage_loop}$^\text{e}$                     & SPDC       & Storage loop              & -      & 27.9  & 0.007 & 91   \\ \hline
\end{tabular}
\begin{tablenotes}[flushleft]
\item[a] Here the source is internal to the memory in a combined system.
\item[b] The enhancement factor $\zeta$ is calculated, not measured.
\item[c] The synchronization rate is estimated from Fig.~4(b) in the main text of Ref.~\cite{2020_Xian_Min_Jin_synchronization}.
\item[d] The visibility was derived by adding the measured background.  
\item[e] Here the storage loop was used not for photon synchronization, but in a time multiplexing scheme to enhance the photon generation probability in a pre-detrmined clock-cycle.
\end{tablenotes}
\end{threeparttable}
\end{table}

\section{Experimental setup} \label{experimental setup}

\subsection{Photon source} \label{photon source setup}
We use the photon source described in Ref.~\cite{Photon_source_paper} and improved in Ref.~\cite{Photon_source_paper_2}. 
The pumping powers used are $300-500$ $\mu$W for pump-780 and $1-6.5$ mW for pump-776, corresponding to a heralded single-photon detection rate from $R_j\sim 50$ kilo-counts per second (kcps) up to $R_j\sim 440$ kcps, in each of the $j=1,2$ channels of the source.
Pump-780 detuning is set to $\Delta_\text{p1}=-1.1$ GHz, and pump-776 is set to two-photon resonance. At the optical depth $\text{OD}=4$ of the photon source, the photons' temporal full-width at half maximum is $0.9-0.98$ ns and slightly increasing with the increase of the pump fields' power. 

The detected heralding efficiency, defined as $\eta_{\text{h}j}=R_j/R_{\text{idler-}j}$, where $R_{\text{idler-}j}$ is the detection rate of idler-$j$ photons, is $\eta_{\text{h}1}=20.9\pm0.6 \%$ and $\eta_{\text{h}2}=15.9\pm0.5 \%$. 
The main reason for the lower heralding efficiency in channel 2 is the higher noise originating from a direct scattering of photons from pump-776 into the idler modes. This scattering can be reduced by rotating the vapor cell holder \cite{Photon_source_paper}, but it is generally different for the two idler modes. Additionally, $R_1$ is higher than $R_2$ by $3.3\pm 0.7\%$, which we attribute to a slightly better phase-matching of signal-1  and idler-1. 

$\eta_{\text{h}2}$ can be easily increased to the value of $\eta_{\text{h}1}$ by employing photon sources from two different vapor cells instead of one multiplexed source from a single vapor cell, and optimizing each vapor cell orientation to reduce pump-776 scattering noise into the corresponding idler mode. This will also enable slightly further increasing the heralding efficiency \cite{Photon_source_paper_2}, as the vapor cells orientation can be chosen to optimize the scattering noise into a single spatial mode, instead of a trade-off of the noise into two different modes.

\subsection{Memory} \label{memory setup}
Our quantum memory is based on the system described in Ref.~\cite{FLAME_2_paper}. We change the vapor cell used in Ref.~\cite{FLAME_2_paper} to a similar one (same size and same type; Precision Glassblowing) with a higher transmission of $96\%$ at 780 nm. The higher transmission is obtained by a different cleaning procedure of immersing the cell into an Acetone bath before inserting it into the setup. The total memory setup transmission, from after the input fiber, is thus increased from $66\%$ in Ref.~\cite{FLAME_2_paper} to $71\%$ in this work. Considering an extra 4\% loss in the input fiber to free-space coupling, the overall memory setup transmission is $T=68\%$. The vapor cell temperature ($ \sim 65~\degree $C) and $\text{OD} = 19\pm 1$ (with optical pumping) are the same as in Ref.~\cite{FLAME_2_paper}.

The detuning of the control field that optimizes the storage efficiency is $\Delta_\text{c}=-50$ MHz, which corresponds to storing and retrieving signal photons on-resonance \cite{FLAME_2_paper}. The control field's peak power at the vapor cell which optimizes the storage efficiency is 1.2 W. 
The atoms are optically pumped within the ground-state manifold to the maximal spin state $|5S_{1/2},F=2,m_F=2\rangle $ using pump and repump fields. The pump beams have an annular shape and are inserted at a slight angle to the optical axis, as described in Ref.~\cite{FLAME_2_paper}.  

We use an auxiliary dressing field that weakly couples the $5D_{5/2}$ level to the $28F_{7/2}$ level with a detuning of $\Delta_\text{d}=-500$ MHz and power of 80 mW. This field counteracts the residual Doppler broadening of the  two-photon transition $|5S_{1/2}\rangle\rightarrow |5P_{3/2}\rangle\rightarrow |5D_{5/2}\rangle$ \cite{Continuous_protection_paper} and prolongs the memory lifetime \cite{FLAME_2_paper}. Using a memory storage time of up to 100 ns as in our demonstration, the rate of synchronized photon pairs increases with the dressing field by $5\pm 2 \%$ compared to without the dressing field. Using longer memory storage times increases the benefit of employing the dressing field.

\subsection{Electronics and timing} \label{electronics}
\begin{figure} 
	\centering
	\includegraphics[width=\columnwidth,trim=2.5cm 15cm 4.0cm 0.0cm,clip=true]{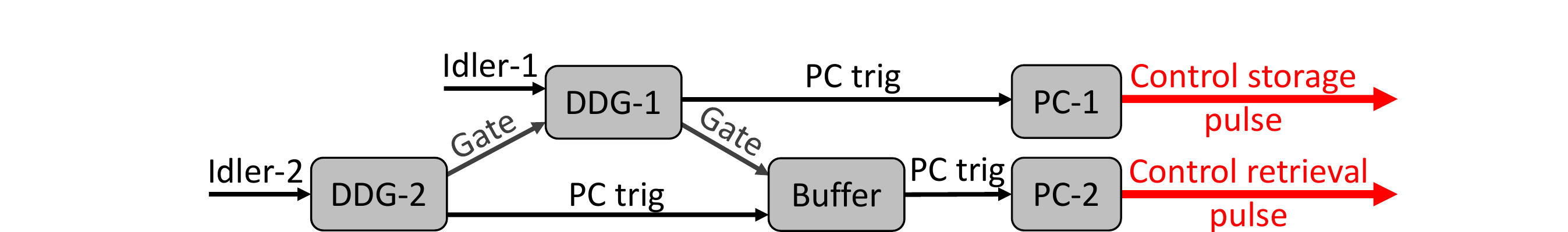}
	\caption{\textbf{Electronic triggering scheme of memory operation.} The PCs are operated only if both idler photons were detected within a $t^*=100$ ns time window. 
	}
	\label{fig:electronics} 
\end{figure}

The maximal average repetition rate of our PCs is limited to $3\times 10^5/$sec and the minimal time between pulse operations is 1.5 $\mu$s. Therefore, we only operate the memory if the two idler photons are detected within a $t^*=100$ ns time window. Figure~\ref{fig:electronics} shows the electronic triggering scheme of the experiment. Upon detection of idler-2 photon, DDG-2 (T564, Highland technology) is triggered and starts the electronic control sequence. Following a 22-ns insertion delay, DDG-2 outputs a gating pulse of length $t^*=100$ ns to DDG-1. After another $\sim 130$ ns, it outputs a trigger to PC-2 which operates the retrieval pulse. This time delay is set to accommodate the electronic delays and memory lifetime, and is controlled with 10-ps resolution. 
DDG-1 is triggered only if an idler-1  detection is received within the $t^*=100$ ns `Gate' pulse from DDG-2. After the 22-ns insertion delay, it sends a trigger pulse with a 15-ns delay (to match the arrival time of signal-1 to the memory) to PC-1, which operates the control storage pulse. It also sends a Gate pulse to the logic Buffer (T860, Highland technology) that transmits the trigger of DDG-2 to PC-2 only if the Gate is on, ensuring PC-2 is operated only when both idler photons were detected within $t^*$. 

The DDGs ignore any input trigger while they output pulses. We use this feature to suppress the PCs triggering in times shorter than 1.5 $\mu$s. For DDG-1, we set a trigger acceptance downtime of $\tau_{\text{d}1}=1.525 \ \mu$s (after each accepted trigger), which ensures that the PCs are not operated within this time. For DDG-2, we set a trigger acceptance downtime of $\tau_{\text{d}2}=260$ ns, which is set slightly longer than required for photon synchronization in order to ensure that the Gate from DDG-1 to the Buffer does not overlap with the trigger pulse from DDG-2  in different memory operation realizations.  
The average duty cycle of the experiment, accounting for the acceptance downtime of the DDGs, is defined as the memory downtime used in Fig.~4 of the main text.

\subsection{Optical fibers and photon detection} \label{optical fibers}
Signal-2 photons generated by the source are coupled to a 60-m-long delay-line optical fiber. This fiber length is, on the one hand, long enough to accommodate the electronics delays and memory lifetime and, on the other hand, short as possible in order to minimize the photon loss ($\le -4$ dB/km) and the forced downtime of the electronics. A slightly longer (shorter) fiber can be chosen to enable working with a longer (shorter) idler photons' detection time window $t^*$. For given generation rates $R_j$, different choices of $t^*$ optimize the synchronized photon-pair rate $\Rsync$. This is due to a trade-off between the probability for two idler photons to be generated within $t^*$, and the average memory efficiency within $t^*$ and the electronics downtime. The model for $\Rsync$ given in Sec.~\ref{coincidence rate model} can be used to find the optimal $t^*$. 

Signal-1 photons generated by the source are coupled to a 25-m-long optical fiber that is coupled with a mating-sleeve connection to a 5-m-long fiber that is input to the memory, or to a different fiber that goes directly to a photon detector. The combined 30-m-long fiber from the source to the memory is used as a delay line to accommodate the electronic latencies and the propagation of the idler photons and control pulses through the optical fibers. The mating sleeve connection enables us to characterize the source without the memory and to input classical light into the memory for characterization and alignment. The coupling efficiency of the mating sleeve connection is $98\pm1 \%$ ($92\pm1 \%$) to the memory fiber (detector fiber).  

We detect the photons using superconducting nanowire single-photon detectors (Quantum Opus) with $91\pm 2 \%$ detection efficiency and $55$-ps detection time jitter. These detectors are sensitive to the polarization of the light fields, and therefore we couple the signal and idler photons from polarization-maintaining fibers to single-mode fibers in polarization controllers (Thorlabs FPC030) using mating-sleeve connectors. The single-mode fibers are then coupled to the detectors. In order to minimize the loss in the mating-sleeve connectors, we use low-loss key-aligned fiber optical patch cables (Oz Optics) with $97.5\pm 1.5 \%$ transmission.

\section{Model for photon coincidence rate} \label{coincidence rate model}
The accidental photon-coincidence rate $R_\text{stoc}$ is given by the signal-2 detection rate $R_2$ multiplied by the probability $p_\text{c}=R_1\delta\text{t}$ that a signal-1 photon is detected within a small time window $\delta\text{t}$. As explained in the main text, we consider $\delta\text{t}=600$ ps. Therefore, 
\begin{equation}
    R_\text{stoc} = R_1R_2\delta\text{t}.
\end{equation}

For the coincidence rate after synchronization $R_\text{sync}$, we first examine the triggering rate $R_{\text{trig-}2}$ of DDG-2. The rate at which idler-2 photons are detected is $R_{\text{idler-}2}=R_2/\eta_{\text{h}2}$. However, the actual triggering rate $R_{\text{trig-}2}= R_{\text{idler-}2}/(1+R_{\text{idler-}2}\tau_{\text{d}2})$ is lower due to the set downtime $\tau_{\text{d}2}$ of DDG-2 after an accepted input trigger. Next, we define the probability $p_{\text{trig-}1}=(R_1/\eta_{\text{h}1})t^*$ that, within the time window $t^*$, idler-1 is detected and triggers DDG-1. The rate of DDG-1 output pulses (which corresponds to the number of synchronization attempts) is also reduced due to its electronics downtime $\tau_{\text{d}1}$. Therefore, the rate of synchronization attempts is 
\begin{equation}
    R_{\text{sync-trials}} = \frac{ R_{\text{trig-}2}p_{\text{trig-}1} }{ 1 + R_{\text{trig-}2}p_{\text{trig-}1} \tau_{\text{d}1} }. 
\end{equation}
Using $R_{\text{sync-trials}}$, we calculate the memory downtime, given by 
\begin{equation}
    \text{Memory downtime} = (R_{\text{trig-}2}-R_{\text{sync-trials}})\tau_{\text{d}2} + R_{\text{sync-trials}}\tau_{\text{d}1}. \label{memory downtime eq}
\end{equation} 
Here we subtract the rate of synchronization attempts from the trigger rate of DDG-2 so as not to double-count the electronics downtime.

The average memory efficiency in the synchronization experiment is $\effbar = \tfrac{1}{t^*}\int_0^{t^*} \eff(t)dt$, which in our experiment is $\effbar = 0.196$ ($\effbar = 0.184$) before (after) correcting for the different transmissions of signal-1 through the two fiber alternatives, \textit{i.e.,} memory fiber or detector fiber. Note that here we neglect the probability of a photon arriving earlier within $t^*$, as it is small for our experimental parameters. The synchronized photon coincidence count rate is finally given by 
\begin{equation}
    R_\text{sync} =  R_{\text{sync-trials}} \eta_{\text{h}1} \eta_{\text{h}2} \effbar,
\end{equation}
and the coincidence enhancement factor is $\zeta = R_\text{sync} / R_\text{stoc}$. 

Figure~4 in the main text shows the calculated $\Rsync$ and $\Rstoc$, exhibiting an excellent agreement with the measured data. We attribute the small reduction of the measured $\Rsync$ compared to the model at high average downtime to a degradation of the amplitude of the control pulses when they are generated within 2 $\mu$s from one another, which slightly reduces $\effbar$. 

\begin{figure} 
	\centering
	\includegraphics[width=\columnwidth,trim=0.0cm 0cm 0.0cm 0.0cm,clip=true]{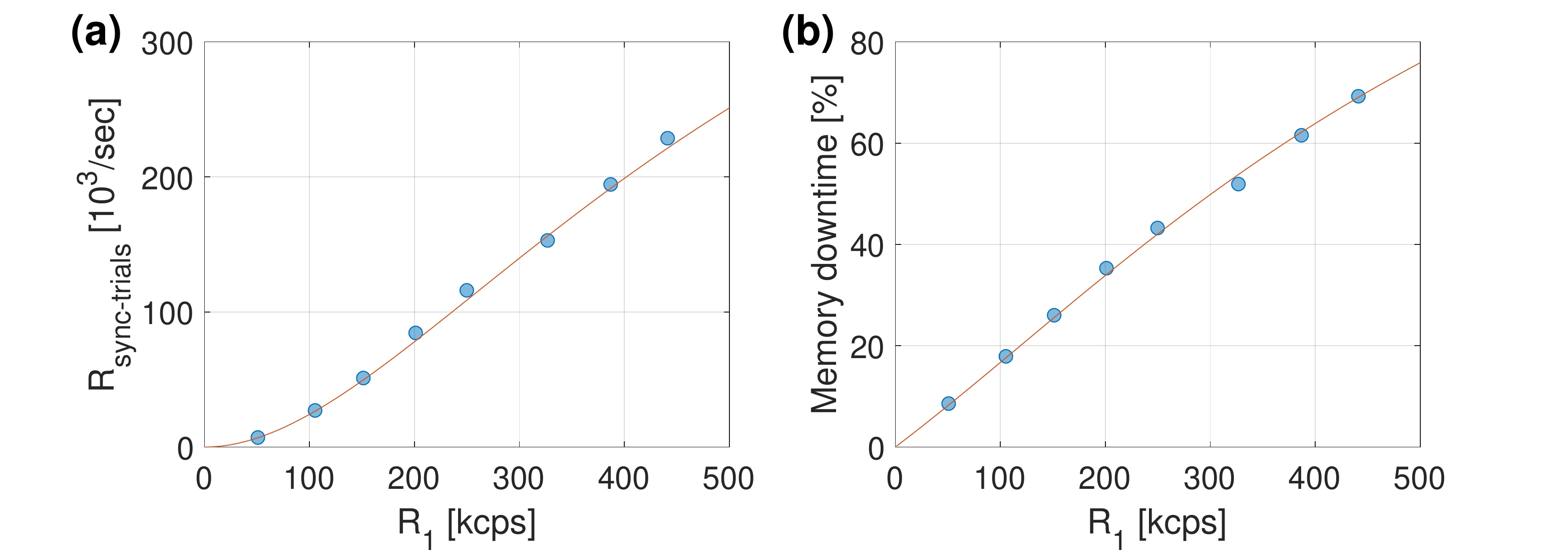}
	\caption{\textbf{Memory operation rate and downtime versus single-photon rate.} 
        (\textbf{a}) The rate of memory operations (synchronization attempts rate) $R_{\text{sync-trials}}$ versus the single photon generation rate $R_1$.
        (\textbf{b}) Memory downtime given by Eq.~\eqref{memory downtime eq}. 
        In (\textbf{a}) and (\textbf{b}), circles are measured data and lines are based on the calculation described in Sec.~\ref{coincidence rate model}. In (\textbf{b}), $R_{\text{trig-}2}$ of the data is inferred from the measured $R_{\text{idler-}2}$ and $\tau_{\text{d}2}$. 
	}
	\label{fig:memory operation rate and downtime} 
\end{figure}

Figure~\ref{fig:memory operation rate and downtime} shows the memory operation rate $R_{\text{sync-trials}}$ and downtime versus $R_1$.  
The model correctly predicts $R_{\text{sync-trials}}$ for all $R_1$ (with $R_2\approx 0.97 R_1$), which further confirms that the slight degradation of the coincidence count rate is due to the reduced control-pulse amplitude.

\section{Model for $\gh$} \label{g2 model}
In the Fock state basis, the heralded single-photon state before the memory is given by
\begin{equation}
    |\psi_\text{1} \rangle \approx \sqrt{1-\eta_\text{h} } |0\rangle + \sqrt{\eta_\text{h} } |1\rangle + \eta_\text{h}\sqrt{\tfrac{1}{2} g_\text{h-1}^{(2)}} |2\rangle,
\end{equation}
where $g_\text{h-1}^{(2)}$ is the conditional, same-time, auto-correlation $\gh(0)$ of signal-1 (before the memory).
Here we assume that $P(1)\gg P(2) \gg P(n>2)$, where $P(n)$ is the probability that $|\psi \rangle $ contains $n$ photons, such that $g_\text{h-1}^{(2)}\approx 2P(2)/[P(1)]^2$ \cite{photon_statistics_review}.

After the memory, the amplitude of the single-photon component reduces to $\sqrt{\eta_\text{h}\eff(t)}$ due to the finite memory efficiency. We assume that the two-photon component is comprised of one heralded single photon (\textit{i.e.} the signal photon) and one noise photon. The signal photon is stored and retrieved from the memory with efficiency $\eff(t)$. The noise photon may arrive from three different sources: 
(1) A (second) signal photon emitted at the same time as the (first) signal photon and stored with efficiency $\eff(t)$; (2) A (second) signal photon emitted during the retrieval time of the memory, which may be partially transmitted through the memory module with transmission $T^\text{during}_\text{retrieval}$; (3) An off-resonant photon scattered from pump-780 into the signal mode at the time of retrieval, transmitted through the memory module with transmission $T_\text{off-res.}$. While for the on-resonance photons, $T^\text{during}_\text{retrieval}<T(1-\sqrt{\eta_\text{int}})\ll 1$, for the off-resonance photons, the transmission $T_\text{off-res.}\approx 0.9T$ is high (the 0.9 correction originates from residual absorption in the Autler-Townes splitting of the atomic absorption line, estimated from a comprehensive simulation of our system). 

With these three sources for the noise photon, the photonic state after storage and retrieval at time $t$ is given by
\begin{equation}
\begin{split}
   |\psi_{1'(t)} \rangle \approx & \sqrt{1-\eta_\text{h}\eff(t) } |0\rangle + \sqrt{\eta_\text{h}\eff(t) } |1\rangle \\ 
   & + \eta_\text{h}\sqrt{\tfrac{1}{2}g_\text{h-1}^{(2)}} \sqrt{(1-\rho) \eta_{\text{e2e}}^2(t) + (1-\rho)\eff(t)T^\text{during}_\text{retrieval}  +\rho\eff(t)T_\text{off-res.} } |2\rangle,
   \end{split}
\end{equation} 
and the conditional auto-correlation after the memory is
\begin{equation} \label{g^2 after the memory}
    g_{\text{h-}1'(t)}^{(2)} = g_\text{h-1}^{(2)} \big[ (1-\rho) +(1-\rho)\frac{T^\text{during}_\text{retrieval}}{\eff(t)} +\rho\frac{T_\text{off-res.}}{\eff(t)} \big].
\end{equation} 
Here $\rho$ is the probability that the noise photon originates from the scattered off-resonant pump-780, which we independently measure.

\begin{figure} 
	\centering
	\includegraphics[width=10cm,trim=0.0cm 0cm 0.0cm 0.0cm,clip=true]{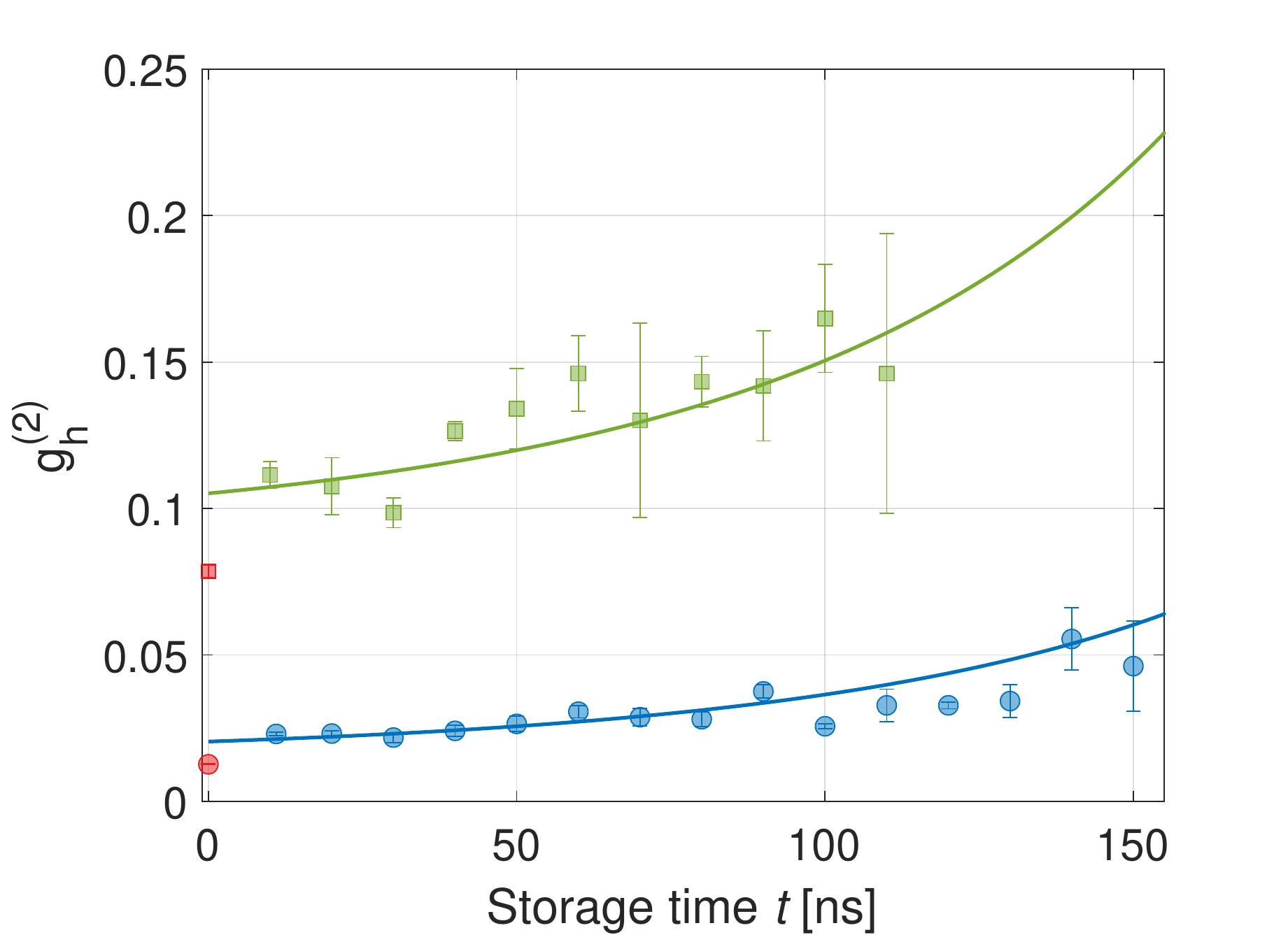}
	\caption{\textbf{$\gh$ for different} $\mathbf{R_1}$. Blue circles are measured data with $R_1\approx 50$ kcps as in Fig.~2(c) of the main text, and green squares are measured data with $R_1\approx 440$ kcps. The lines are a fit to Eq.~\eqref{g^2 after the memory} with the same $T^\text{during}_\text{retrieval}$ for both data sets. Red data points are the $\gh$ of the unstored photons for comparison.
	}
	\label{fig:g_2} 
\end{figure}

We fit Eq.~\eqref{g^2 after the memory} to the measured $\gh$ after the memory with $T^\text{during}_\text{retrieval}$ as the only fit parameter. As shown in Fig.~\ref{fig:g_2}, we simultaneously fit the model to $\gh$ with $R_1\approx50$ kcps [as shown in Fig.~2(c) in the main text] and with $R_1\approx 440$ kcps.
The probability $\rho$ is independently measured to be $\rho=0.35$ ($\rho=0.1$) for $R_1\approx50$ ($R_1\approx440$). Here the difference arises from different ratios between the powers of pump-780 and pump-776. We extract from the fit $T^\text{during}_\text{retrieval}=0.1T$.

\section{Data analysis} \label{data analysis}
Here we provide details on the data analysis used to compile the figures of the main text. 
Throughout the analysis, we consider a heralded single-photon detection if a signal photon is detected within 3.5-ns of the idler photon, or within a similar time window shifted due to the memory operation, as shown in Fig.~2(a) in the main text. This window accounts for $>95\%$ of the photon-pulse energy. We consider only detected photon events and do not subtract any background.

The memory end-to-end efficiency is calculated by integrating the photons' energy directly after the photon source and after storage and retrieval in the memory. Due to the lower transmission of signal-1 in the fiber coupling to the detector (see Sec.~\ref{optical fibers}), we multiply the measured memory efficiency and enhancement factor $\zeta$ by $0.92/0.98\approx 0.94$, so as not to overestimate our memory performance. 
The error bars in Figs.~2(c), 3(b), and 4(a) are calculated from the standard deviation of the mean (STDM) of repeated measurements. 
Due to a $\sim 3\%$ uncertainty in the fiber transmissions and detection efficiency, 
we calculate the error bars of $\eff$ as
\begin{equation}
    [\Delta\eff](t) = \sqrt{ [\text{STDM(t)}]^2 + [0.03\eff(t)]^2  },
\end{equation}
with a corresponding expression for the error bars of $\zeta$. 

The signal photon auto-correlation $\gh$, conditioned on an idler photon detection, is measured using a Hanburry Brown and Twiss setup, where the signal photon is split in a fiber beam splitter and detected in the output modes $\text{s}_\text{a}$ and $\text{s}_\text{b}$. The conditional auto-correlation is given by $\gh = R_{\text{i-s}_\text{a}\text{-s}_\text{b}} R_\text{idler} /(R_{\text{i-s}_\text{a}}R_{\text{i-s}_\text{b}}) $ \cite{1986_Grangier_cascade_source}, where $R_{\text{i-s}_\text{a}\text{-s}_\text{b}}$ is the three-photon coincidence count rate in a 3.5 ns time window, and $R_{\text{i-s}_\text{a}}$ and $R_{\text{i-s}_\text{b}}$ are the two-photon coincidence count rates in the same time window.

We measure the indistinguishability of signal-1 and signal-2 using a HOM interference measurement \cite{HOM_1987}, with the two photons entering a symmetric fiber BS from different input ports. 
For the HOM measurement without synchronization, the time axis in Fig.~3(b) of the main text is given by the difference in detection times of the idler photons $\Delta t = t_{\text{idler-}1}-t_{\text{idler-}2}$. We consider a double-heralded-single-photon event if $t_{\text{s}_\text{a}}-t_{\text{idler-}1}$ \emph{and} $t_{\text{s}_\text{b}}-t_{\text{idler-}2}$ are both detected within the 3.5-ns window shown in Fig.~2(a) of the main text, \emph{or} $t_{\text{s}_\text{a}}-t_{\text{idler-}2}$ \emph{and} $t_{\text{s}_\text{b}}-t_{\text{idler-}1}$ are detected within the time window. Here $t_{\text{s}_\text{a}}$ and $t_{\text{s}_\text{b}}$ are the detection times of the signal photons at the output of the beam splitter in modes $\text{s}_\text{a}$ and $\text{s}_\text{b}$, respectively.   

For the synchronized HOM measurement, $\Delta t$ is defined by the different controlled retrieval times of signal-1$'$ from the memory. 
Here we consider a double-heralded-single-photon event if $t_{\text{s}_\text{a}}-t_{\text{i2}}$ \emph{and} $t_{\text{s}_\text{b}}-(t_{\text{i2}}+\Delta t)$ were both detected within the 3.5-ns time window, \emph{or} $t_{\text{s}_\text{a}}-(t_{\text{i2}}+\Delta t)$ \emph{and} $t_{\text{s}_\text{b}}-t_{\text{i2}}$ were detected within the time window. Here $t_\text{i2}$ is idler-2 detection time, conditioned on the memory operation, which ensures that idler-1 (now with uncorrelated timing to signal-1) was also detected. 

For Fig.~4(a) in the main text, a double-heralded-single-photon coincidence for $R_\text{stoc}$ is considered if $t_\text{signal-1}-t_\text{idler-1}$ \emph{and} $t_\text{signal-2}-t_\text{idler-2}$ are both detected within a 3.5-ns window, \emph{and} $|t_\text{idler-1}-t_\text{idler-2}|\le 300$ ps. 
We consider a double-heralded-single-photon coincidence for $R_\text{sync}$ if $t_{\text{signal-1}'}-t_\text{i2}$ \emph{and} $t_\text{signal-2}-t_\text{i2}$ are both detected within a 3.5-ns window, where $t_{i2}$ is defined above.

\begin{figure} 
	\centering
	\includegraphics[width=10cm,trim=0.0cm 0cm 0.0cm 0.0cm,clip=true]{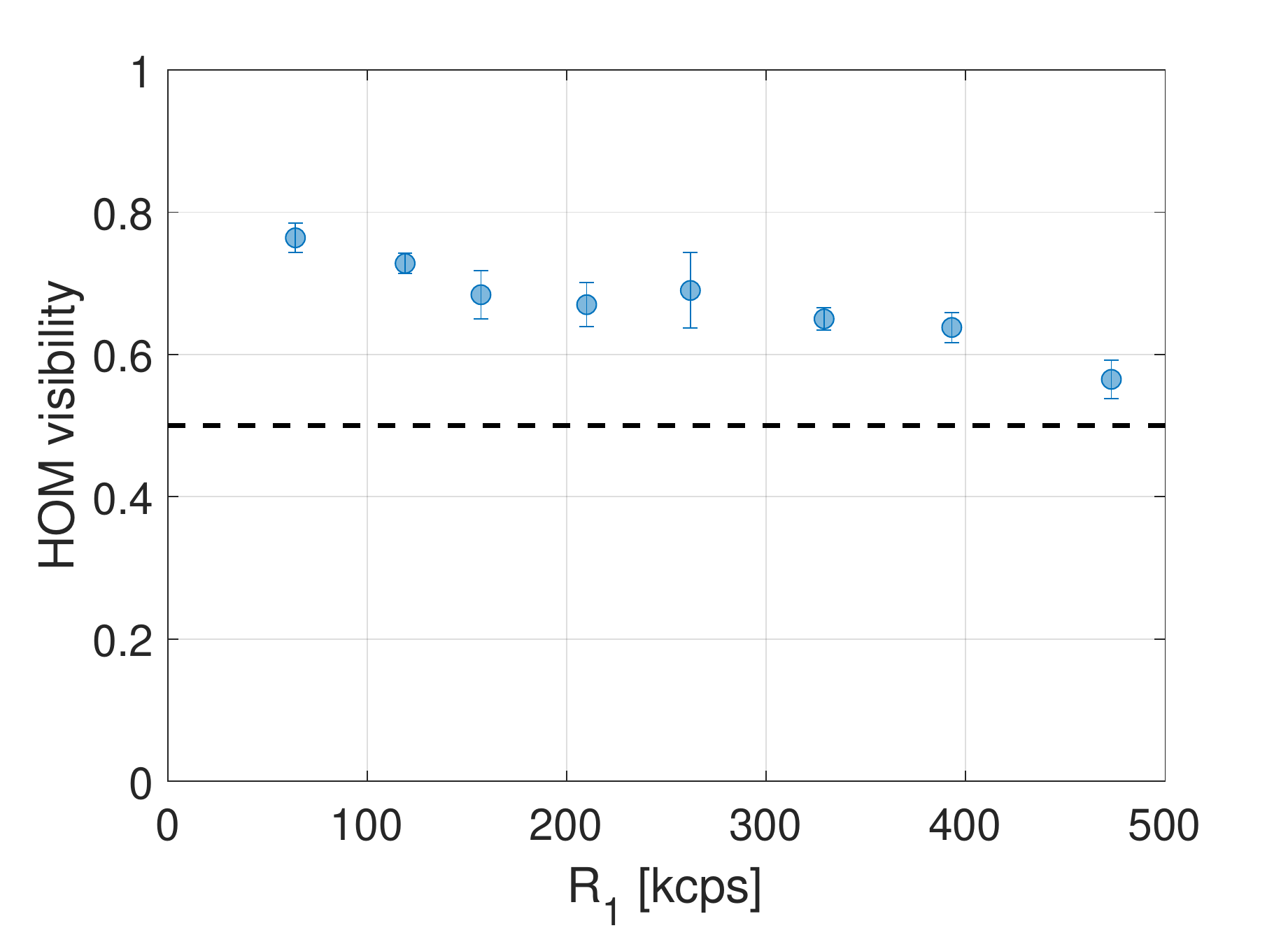}
	\caption{\textbf{HOM visibility $V_\text{sync}$ versus $R_1$.} 
        The dashed line represents the non-classical bound. In this data set, $R_1$ is inferred from the DDGs triggering rate.
	}
	\label{fig:HOM vis} 
\end{figure}

\section{Additional experimental data} \label{Additional experimental data}
Figure~\ref{fig:HOM vis} shows the HOM interference visibility $V_\text{sync}$ versus $R_1$. As $R_1$ increases, $V_\text{sync}$ decreases due to the increase in the multi-photon component $\gh$. $V_\text{sync}$ is reduced compared to $V_\text{stoc}$ partially due to the increase of $\gh$ and partially due to the temporal distortion of the retrieved photon. To evaluate the latter, we calculate the temporal overlap of signal-1$'$ and signal-2 \cite{2019_Du_cold_memory} $I=|\int \sqrt{G_{\text{s}{1'}\text{i}2}(\tau) G_{\text{s}{2}\text{i}2}(\tau)}d\tau|^2 / M$, where $M=[\int G_{\text{s}{1'}\text{i}2}(\tau) d\tau \int G_{\text{s}{2}\text{i}2}(\tau) d\tau]$,
using the red and solid-blue profiles in Fig.~3(a) in the main text. We find $I=91\%$, making it the dominant source of the visibility reduction at low $R_1$, while the increased $\gh$ is the dominant source of the visibility reduction at high $R_1$.
For all data points presented in the paper, $V_\text{sync}>0.5$ exceeds the classical bound.



\end{document}